\documentclass[12pt]{article}

\usepackage{geometry}
\usepackage{cite}
\usepackage{paralist}
\usepackage{graphicx}
\usepackage{subfigure}
\usepackage{amssymb}
\usepackage{amsmath}
\allowdisplaybreaks[1]

\usepackage{amsthm}

\makeatletter
\def\@endtheorem{\endtrivlist}
\makeatother

\newcounter{Rule}
\newenvironment{brule}{\refstepcounter{Rule}\par\smallskip\noindent
\textbf{(\arabic{Rule})}\quad}{}
\newenvironment{rrule}{\refstepcounter{Rule}\par\smallskip\noindent
\textbf{(\arabic{Rule})}\quad}{}
\newcommand{\currentrule}{\arabic{Rule}}

\newcommand{\edge}[2]{#1#2}
\newcommand{\Fmin}[1]{\mathcal{F}(#1)}

\begin{document}

\title{An algorithm for destroying claws and diamonds}
\author{Dekel Tsur%
\thanks{Ben-Gurion University of the Negev.
Email: \texttt{dekelts@cs.bgu.ac.il}}}
\date{}
\maketitle

\begin{abstract}
In the \textsc{\{Claw,Diamond\}-Free Edge Deletion} problem the input is
a graph $G$ and an integer $k$, and the goal is to decide whether there
is a set of edges of size at most $k$ such that removing the edges of the
set from $G$ results a graph that does not contain an induced claw or diamond.
In this paper we give an algorithm for this problem
whose running time is $O^*(3.562^k)$.
\end{abstract}


\section{Introduction}

A \emph{claw} is a graph with four vertices such that one vertex,
called the \emph{center}, is adjacent to the other three vertices of the graph,
and there are no edges between the non-center vertices.
A \emph{diamond} is a graph with four vertices and five edges.
In other words, there is an edge between every pair of vertices except one pair.
In the \textsc{\{Claw,Diamond\}-Free Edge Deletion} problem the input is
a graph $G$ and an integer $k$, and the goal is to decide whether there
is a set of edges of size at most $k$ such that removing the edges of the
set from $G$ results a graph that does not contain an induced claw or diamond.
This problem was introduced by Cygan et al.~\cite{cygan2017polynomial}.
The importance of this problem is due to its connection to the
\textsc{Claw-Free Edge Deletion} problem.
It is currently an open problem whether the latter problem has a polynomial
kernel, and a polynomial kernel for
\textsc{\{Claw,Diamond\}-Free Edge Deletion} may be a first step towards
a polynomail kernel for \textsc{Claw-Free Edge Deletion} (see the discussion
in~\cite{cygan2017polynomial}).

Cygan et al.~\cite{cygan2017polynomial} showed that the
\textsc{\{Claw,Diamond\}-Free Edge Deletion} problem is NP-hard.
Moreover, the problem has no subexpontial-time algorithm, unless
the exponential time hypothesis fails.
The problem has a simple $O^*(5^k)$-time algorithm~\cite{cai1996fixed}.
Li et al.~\cite{li2017improved} gave an $O(3.792^k)$-time algorith.
In this paper we give an algorithm for whose running time is $O^*(3.562^k)$.

\paragraph{Preleminaries}
For a set $S$ of vertices in a graph $G$, $G[S]$ is the subgraph
of $G$ induced by $S$ (namely, $G[S]=(S,E\cap (S\times S))$).
For a set $F$ of edges, $G-F$ is the graph obtained from $G$ by deleting
the edges of $F$.
For a set $\mathcal{H}$ of graphs and a graph $G$, the graph $G$
is called \emph{$\mathcal{H}$-free} if $G$ does not contain an induced subgraph
which is isomorphic to a graph in $\mathcal{H}$.
A set $F$ of edges is called a \emph{deletion set} if $G-F$ is 
\{claw,diamond\}-free.

\section{The algorithm}

The algorithm is a branching algorithm (cf.~\cite{cygan2015parameterized}).
Given an instance $(G,k)$, the algorithm applies the first applicable rule from
the rules below.
When we say that the algorithm \emph{branches} on sets $F_1,\ldots,F_p$, we mean
that the algorithm is called recursively on the instances
$(G-F_1,k-|F_1|),\ldots,(G-F_p,k-|F_p|)$.
For a graph $H$, let $\Fmin{H}$ be a set containing every inclusion minimal
deletion set of $H$.

\begin{rrule}
If $k < 0$ return `no'.
\end{rrule}

\begin{rrule}
If $G$ is \{Claw,Diamond\}-free, return `yes'.\label{rule:terminate-yes}
\end{rrule}

\begin{brule}
If there is a set $X$ of vertices that induces a claw,
branch on every set in $\Fmin{G[X]}$.\label{rule:claw}
\end{brule}

Rule~(\currentrule) is clearly safe.
The set $\Fmin{G[X]}$ consists of three sets of size~1, where each set consists
of one edge of $G[X]$.
Therefore, the branching vector of Rule~(\currentrule) is $(1,1,1)$
and the branching number is 3.

For the following rules, we have that the graph is claw-free since
Rule~(\ref{rule:claw}) cannot be applied.
Additionally, due to Rule~(\ref{rule:terminate-yes}), the graph contains
at leas one induced diamond.
Assume that $\{a,b,c,d\}$ induces a diamond,
where $b,d$ are not adjacent.

\begin{brule}
If $N(a) \setminus \{c\} = N(c)\setminus \{a\}$, branch on
$\{\edge{a}{c}\}$, $\{\edge{a}{b}\}$, and $\{\edge{a}{d}\}$.\label{rule:twins}
\end{brule}

To prove the correctness of Rule~(\currentrule), note that
$\Fmin{G[a,b,c,d]} = \{  \{\edge{a}{c}\}, \{\edge{a}{b}\},\allowbreak
\{\edge{a}{d}\}, \{\edge{c}{b}\}, \{\edge{c}{d}\}\}$.
Rule~(\currentrule) branches on the first three sets of $\Fmin{G[a,b,c,d]}$.
From the assumption $N(a) \setminus \{c\} = N(c)\setminus \{a\}$ we have that
the graph $G-\{\edge{c}{b}\}$ is isomorphic to the graph
$G-\{\edge{a}{b}\}$.
Therefore, the algorithm does not need to branch on $\{\edge{c}{b}\}$.
Similarly, the graph $G-\{\edge{c}{d}\}$ is isomorphic to the graph
$G-\{\edge{a}{d}\}$.

The branching vector of Rule~(\currentrule) is $(1,1,1)$ and the branching
number is 3.

For the following rules we have that
$N(a) \setminus \{c\} \neq N(c)\setminus \{a\}$.
Let $t$ be a vertex that is adjacent to exactly one vertex from $a,c$.
Without loss of generality, assume that $t$ is adjacent to $a$ and not adjacent
to $c$.

\begin{brule}
If there is a vertex $s \neq t$ that is adjacent to exactly one vertex from
$a,c$, branch on every set in $\Fmin{G[\{a,b,c,d,s,t\}]}$.\label{rule:s+t}
\end{brule}

Rule~(\currentrule) is clearly safe.
To compute the branching number of this rule, we need to consider all
possible cases of the graph $G[\{a,b,c,d,s,t\}]$.
\begin{itemize}
\item
The vertex $t$ must be adjacent to at least one vertex from $b,d$, otherwise
$\{a,b,d,t\}$ induces a claw.
Due to symmetry, it suffices to only consider the cases
$N(t) \cap \{b,d\} = \{b\}$ and $N(t) \cap \{b,d\} = \{b,d\}$.
\item
Using the same argument as above, the vertex $s$ is adjacent to at least
one vertex from $b,d$.
Therefore, $N(s) \cap \{a,b,c,d\}$ is either $\{a,b\}$, $\{a,d\}$, $\{a,b,d\}$,
$\{c,b\}$, $\{c,d\}$, or $\{c,b,d\}$.
\item
If $s$ is adjacent to $c$, the vertices $s,t$ can be either adjacent or
non-adjacent.
If $s$ is adjacent to $a$ then $s,t$ are adjacent, otherwise $\{a,c,s,t\}$
induces a claw.
\end{itemize}
Therefore, there are 18 possible cases for the graph $G[\{a,b,c,d,s,t\}]$.
For each case, we used a Python script to compute $\Fmin{G[\{a,b,c,d,s,t\}]}$.
The case with largest branching number is when $N(t)\cap \{b,d\} =\{b\}$ and
$N(s) \cap \{a,b,c,d\} = \{a,b\}$.
In this case, the branching vector is $(1,1,2,2,2,2,3,3,3,3,3)$ and the
branching number is at most 3.533.

If Rules~(1)--(\currentrule) cannot be applied,
$t$ is the only vertex in $G$ that is adjacent to exactly one vertex from $a,c$.
As shown above, $t$ is adjacent to at least one vertex from $b,d$.
Without loss of generality assume that $t$ is adjacent to $b$.

\begin{brule}
If $t$ is not adjacent to $d$, branch on every set in
$\Fmin{G[\{a,b,c,d,t\}]}$ except $\{\edge{a}{t},\edge{c}{d}\}$.\label{rule:t-1}
\end{brule}

We have that $G-\{\edge{a}{t},\edge{c}{d}\}$ is isomorphic to
$G-\{\edge{a}{t},\edge{a}{d}\}$ and
$\{\edge{a}{t},\edge{a}{d}\} \in \Fmin{G[\{a,b,c,d,t\}]}$.
Therefore, Rule~(\currentrule) is safe.
Since
$\Fmin{G[\{a,b,c,d,t\}]} =  \{ \{\edge{a}{b}\}, \{\edge{b}{c}\},
\{\edge{a}{c}\}, \{\edge{a}{t},\edge{a}{d}\},\allowbreak
\{\edge{a}{t},\edge{c}{d}\},\{\edge{b}{t},\edge{a}{c}\}\}$,
the branching vector of Rule~(\currentrule) is $(1,1,1,2,2)$
and the branching number is at most 3.562.

\begin{brule}
Otherwise (namely, if $t$ is adjacent to $b$), branch on every set in
$\Fmin{G[\{a,b,c,d,t\}]}$ except $\{\edge{a}{t},\edge{c}{b}\}$
and $\{\edge{a}{t},\edge{c}{d}\}$.
\end{brule}

As before, the safeness of Rule~(\currentrule) follows from symmetry:
The graph $G-\{\edge{a}{t},\edge{c}{b}\}$ is isomorphic to
$G-\{\edge{a}{t},\edge{a}{b}\}$ and
$\{\edge{a}{t},\edge{a}{b}\} \in \Fmin{G[\{a,b,c,d,t\}]}$.
Additionally, the graph $G-\{\edge{a}{t},\edge{c}{d}\}$ is isomorphic to
$G-\{\edge{a}{t},\edge{a}{d}\}$ and
$\{\edge{a}{t},\edge{a}{d}\} \in \Fmin{G[\{a,b,c,d,t\}]}$.
The branching vector of Rule~(\currentrule) is $(2,2,2,2,2,2,2,2,2,2,2,2)$
and the branching number is at most 3.465.

The rule with the largest branching number is Rule~(\ref{rule:t-1}).
Therefore, the running time of the algorithm is $O^*(3.562^k)$.

\bibliographystyle{abbrv}
\bibliography{claw-diamond}

\end{document}